# Fano collective resonance as complex mode in a two-dimensional planar metasurface of plasmonic nanoparticles

Salvatore Campione[1,$,#], Domenico de Ceglia[2], Caner Guclu[1], Maria A. Vincenti[2], Michael Scalora[3], and Filippo Capolino[1,*]

[1]Department of Electrical Engineering and Computer Science, University of California Irvine, Irvine CA 92697 USA
[2]National Research Council – AMRDEC, Charles M. Bowden Research Laboratory, Redstone Arsenal, AL, 35898, USA
[3]Charles M. Bowden Research Laboratory, AMRDEC, US Army RDECOM, Redstone Arsenal, AL, 35898, USA
[$]Current address: Center for Integrated Nanotechnologies (CINT), Sandia National Laboratories, P.O. Box 5800, Albuquerque, NM, 87185, USA
[#]sncampi@sandia.gov ; [*]f.capolino@uci.edu

*Abstract*—**Fano resonances are features in transmissivity/reflectivity/absorption that owe their origin to the interaction between a bright resonance and a dark (i.e., sub-radiant) narrower resonance, and may emerge in the optical properties of planar two-dimensional (2D) periodic arrays (metasurfaces) of plasmonic nanoparticles. In this Letter, we provide a thorough assessment of their nature for the general case of normal and oblique plane wave incidence, highlighting when a Fano resonance is affected by the mutual coupling in an array and its capability to support free modal solutions. We analyze the representative case of a metasurface of plasmonic nanoshells at ultraviolet frequencies and compute its absorption under TE- and TM-polarized, oblique plane-wave incidence. In particular, we find that plasmonic metasurfaces display two distinct types of resonances observable as absorption peaks: one is related to the Mie, dipolar resonance of each nanoparticle; the other is due to the *forced excitation* of free modes with small attenuation constant, usually found at oblique incidence. The latter is thus an *array-induced collective Fano resonance*. This realization opens up to manifold flexible designs at optical frequencies mixing individual and collective resonances. We explain the physical origin of such Fano resonances using the modal analysis, which allows to calculate the *free modes* with complex wavenumber supported by the metasurface. We define equivalent array dipolar polarizabilities that are directly related to the absorption physics at oblique incidence and show a direct dependence between array modal phase and attenuation constant and Fano resonances. We thus provide a more complete picture of Fano resonances that may lead to the design of filters, energy-harvesting devices, photodetectors, and sensors at ultraviolet frequencies. Similar resonances may be also extended to the visible range with an appropriate choice of geometries and materials.**

The optical properties of plasmonic nanoparticles (e.g. scattering and absorption cross sections) may exhibit features sometimes referred to as *Fano resonances* [1-2], arising from the interaction between a broad, bright resonance and a narrow, dark one. The literature abounds with references to Fano resonances in plasmonic nanostructures [3-17]. Fano resonances have been proposed for ultrasensitive spectroscopy [10] and artificial magnetism at optical frequencies [15, 18-19]. More recently it has been shown that a normal-incidence analysis of two-dimensional (2D) periodic arrays of plasmonic nanoshells [13] may be employed to realize Fano-like resonances at ultraviolet frequencies, even though they do not exhibit high quality factors. In this Letter, we extend the analysis to arbitrary angles of incidence and expand the class of Fano resonances that are supported by the metasurface of plasmonic nanoparticles. In particular, we define when a Fano resonance is or is not observed in the metasurface's optical properties. These Fano resonances can be narrow if well engineered. However, our main purpose is to explain their physical origin by interrelating them to the free modes supported by the metasurface itself, rather than showing yet another result of high-quality factor Fano resonance. To the authors' knowledge, this analysis has not been done before in the context of Fano resonances; a similar approach has been employed recently to model the properties of "near-zero-index metasurfaces" [20]. As discussed in [13], at normal incidence the nanoshells' Mie dipolar resonances may be observed as absorption peaks, for example. We show here that absorption spectra may be dramatically changed at oblique incidence, and that under this condition other *array-induced* collective resonances appear, as explained next. These resonances provide design flexibility to achieve enhanced, narrow optical properties in plasmonic metasurfaces. The complex modal analysis that we present here clarifies the nature of other resonant mechanisms (different from Mie resonances) that involve the whole array and usually appear at oblique incidence.

For this reason we first show the absorption spectra at normal and oblique incidence of the metasurface of plasmonic nanoshells (Fig. 1) analyzed in [13], highlighting the differences. We consider here arrays in homogeneous host media to better focus on physical explanations, though experimental realizations would in general involve the presence of a substrate. The latter would affect the optical properties, though the array can be designed in order to compensate for substrate effects or even to improve the Fano-resonances' selectivity (see, e.g., [15]). We analyze both transverse electric (TE) and transverse magnetic (TM), obliquely-incident plane waves. We then investigate the free modes with complex wavenumber that can be supported by the metasurface and analyze the absorption versus frequency and incidence angle. Some of these spectral features, i.e., Fano resonances, are attributable to *forced excitation* of *free modes* supported by the array at oblique incidence, as proposed in [21-22]. To stress this concept, we further define an equivalent *array polarizability* as done in [23] that accounts for array effects and dramatically affects the Mie dipolar polarizabilities at oblique incidence, which is a typical signature of Fano resonances in arrays of nanoparticles.

We consider a metasurface of plasmonic nanoshells (in what follows a *nanoshell* refers to a nanoparticle with sapphire core and aluminum shell) located on the *x*-*y* plane as in Fig. 1, immersed in a homogeneous background with relative permittivity $\varepsilon_h$. Despite the particular example chosen here, the illustrated phenomena and discussion are general. The nanoshells are placed at positions $\mathbf{r}_{mn} = \mathbf{r}_{00} + \mathbf{d}_{mn}$, where $\mathbf{d}_{mn} = ma\hat{\mathbf{x}} + nb\hat{\mathbf{y}}$, with $m,\ n = 0,\ \pm 1,\ \pm 2,\ ...$, $\mathbf{r}_{00} = x_{00}\hat{\mathbf{x}} + y_{00}\hat{\mathbf{y}} + z_{00}\hat{\mathbf{z}}$, and $a$ and $b$ are the periods along the $x$ and $y$ directions, respectively [24-26].

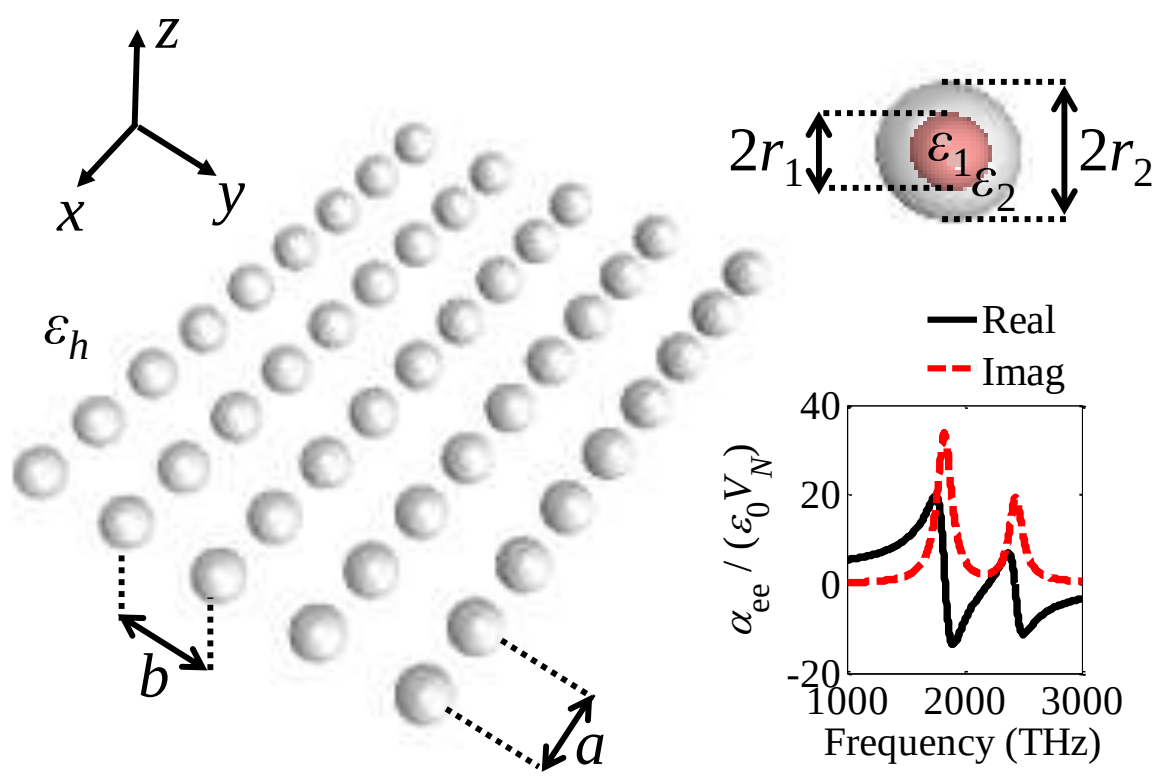

Fig. 1. (Color online) Sketch of a 2D periodic array of plasmonic nanoshells as in the inset, with periods *a* and *b* along *x* and *y* directions, embedded in a homogeneous medium with relative permittivity $\varepsilon_h$. The nanoshell's core has relative permittivity $\varepsilon_1$ and radius $r_1$ and its shell has relative permittivity $\varepsilon_2$ and outer radius $r_2$. The inset also shows the real and imaginary parts of the electric polarizability $\alpha_{ee}$ (normalized by the free space absolute permittivity $\varepsilon_0$ and the nanoshell volume $V_N = 4\pi r_2^3/3$) versus frequency for a nanoshell with a sapphire core with $r_1 = 3\,\mathrm{nm}$ and an aluminum shell with $r_2 = 8.5\,\mathrm{nm}$. Material parameters for core and shell are from Palik's Handbook [27], and $\varepsilon_h = 1$. We consider *a* = *b* = 21.2 nm.

We then investigate the response of the array in Fig. 1 to an incident plane wave and its capability to support free electromagnetic modes (i.e., in absence of excitation). In both cases the field is periodic except for a phase shift described by a Bloch wavevector $\mathbf{k}_\mathrm{B} = k_x^{\mathrm{mode}}\hat{\mathbf{x}} + k_y^{\mathrm{mode}}\hat{\mathbf{y}}$ that, for the modal guided wave, is in general complex and accounts also for decay. Consequently, due to its subwavelength size, the *mn*-th nanoshell is equivalently represented by the electric dipole moment $\mathbf{p}_{mn} = \mathbf{p}_{00}\exp(i\mathbf{k}_\mathrm{B}\cdot\mathbf{d}_{mn})$, where $\mathbf{p}_{00} = \alpha_{ee}\mathbf{E}^{\mathrm{loc}}(\mathbf{r}_{00})$ represents the electric dipole moment of the reference nanoshell according to the single dipole approximation (SDA) [24-26, 28]; $\alpha_{ee}$ is the electric polarizability of the nanoshell (for which we employ the expression in [28-29] derived by the Mie theory); and $\mathbf{E}^{\mathrm{loc}}$ is the local electric field produced by all the nanoshells of the array except the 00-th reference one, plus any externally incident field, if present. This dipolar representation is generally considered a good approximation when the electric dipole term dominates the scattered-field multipole expansion, specifically when the nanoshell dimensions are much smaller than the operating wavelength, and when the periods $a, b \geq 3r_2$. However, accurate results may still be obtained even for smaller periods as shown below through comparison with full-wave simulations based on the finite element method (COMSOL). Real and imaginary parts of $\alpha_{ee}$ (normalized by the free space absolute permittivity $\varepsilon_0$ and the nanoshell volume $V_N = 4\pi r_2^3/3$) versus frequency are reported in the inset of Fig. 1. In agreement with [13], we observe two resonances from an isolated nanoshell with dimensions and materials described in the caption. We refer to these resonances as Mie resonances.

The local electric field acting on the reference nanoshell at position $\mathbf{r}_{00}$ in the array is given by

$$\mathbf{E}^{\mathrm{loc}}(\mathbf{r}_{00},\mathbf{k}_\mathrm{B}) = \mathbf{E}^{\mathrm{inc}}(\mathbf{r}_{00}) + \underline{\breve{\mathbf{G}}}^{\infty}(\mathbf{r}_{00},\mathbf{r}_{00},\mathbf{k}_\mathrm{B})\cdot\mathbf{p}_{00}, \qquad (1)$$

where $\mathbf{E}^{\mathrm{inc}}$ is the incident electric field, and $\underline{\breve{\mathbf{G}}}^{\infty}(\mathbf{r}_{00},\mathbf{r}_{00},\mathbf{k}_\mathrm{B})$ accounts for all the mutual couplings between all *mn*-indexed electric dipoles and $\mathbf{p}_{00}$ and therefore it is not singular at $\mathbf{r} = \mathbf{r}_{00}$. The term $\underline{\breve{\mathbf{G}}}^{\infty}$ is the regularized periodic dyadic Green's function (GF) as defined in [25-26]. Substituting the expression for the local field in Eq. (1) into the electric dipole moment expression, and assuming the nanoshells be polarized along a predetermined direction $v = x, y,$ or $z$, one obtains $\mathbf{p}_{00} = \hat{\mathbf{v}}\,p_{00}$, where $p_{00}$ satisfies the scalar equation $p_{00} = \alpha_{ee}E_v^{\mathrm{inc}}(\mathbf{r}_{00}) + \alpha_{ee}\breve{G}_{vv}^{\infty}(\mathbf{r}_{00},\mathbf{r}_{00},\mathbf{k}_\mathrm{B})\,p_{00}$, with $\breve{G}_{vv}^{\infty} = \hat{\mathbf{v}}\cdot\underline{\breve{\mathbf{G}}}^{\infty}\cdot\hat{\mathbf{v}}$ being the proper diagonal component of the 2D periodic regularized dyadic GF and $E_v^{\mathrm{inc}} = \hat{\mathbf{v}}\cdot\mathbf{E}^{\mathrm{inc}}$. This leads to the scalar equation

$$\left[1 - \alpha_{ee}\breve{G}_{vv}^{\infty}(\mathbf{r}_{00},\mathbf{r}_{00},\mathbf{k}_\mathrm{B})\right]p_{00} = \alpha_{ee}E_v^{\mathrm{inc}}(\mathbf{r}_{00}). \qquad (2)$$

Note that, in general, more complicated structures comprising for example a multilayered environment or anisotropic nanoparticles involve the solution of a matrix system to determine the dipole moment vector $\mathbf{p}_{00}$. After solving Eq. (2), we are then able to calculate the metasurface absorption as $A = 1 - |R|^2 - |T|^2$, where $|R|^2$ is the reflectivity and $|T|^2$ is the transmissivity. Note that for this case the Bloch wavevector $\mathbf{k}_\mathrm{B}$ is purely real and dictated by the transverse-to-z component of the wavevector of the incident plane wave $\mathbf{E}^{\text{inc}}$. We plot the absorption $A$ versus frequency in Fig. 2 for (i) a normal incident plane wave, and (ii) TE- and TM-polarized plane waves incident at 30 degrees. In the same figure, we also show the remarkable agreement between SDA and full-wave simulations (minor dissimilarities are observed because $a,b<3r_2$). The dipolar, Mie resonances of the nanoshells shown in the inset in Fig. 1 induce polarization-independent absorption peaks at ~1750 THz and ~2400 THz. At normal incidence, only these two features dominate the metasurface response, showing that no Fano resonances are visible at normal incidence in this case. However, the situation changes considerably at oblique incidence: in addition to the Mie resonances of the single nanoshell at ~1750 THz and ~2400 THz, the array exhibits two other resonances at ~2100 THz and ~2800 THz. Here we clarify that the latter are Fano resonances arising from the interaction with the free modes supported by the metasurface. The understanding of this phenomenon in general could facilitate the engineering of Fano resonances appearing in arrays' optical properties along with their line widths.

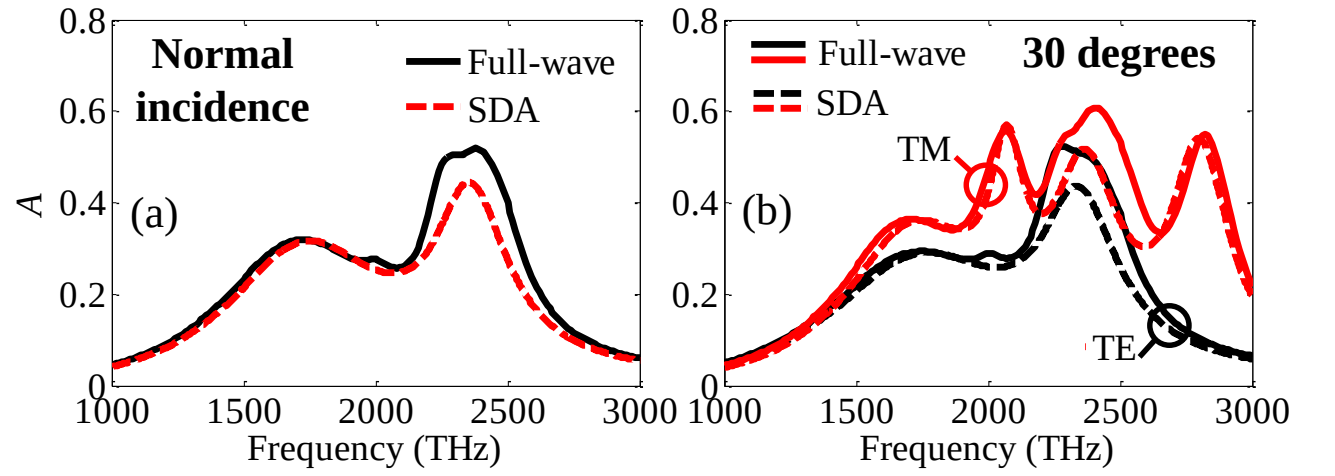

Fig. 2. (Color online) (a) Absorption versus frequency for a normal incident (TE- or TM-polarized) plane wave. (b) As in (a), for 30-degree incidence angle. Single dipole approximation (SDA) and full-wave simulation results are in good agreement. The absorption spectrum is dramatically modified at oblique incidence due to the appearance of array-dependent Fano resonances.

We now continue our analysis by computing the free modes with complex wavenumber supported by the metasurface. Mode analysis is performed by computing the zeroes of the homogeneous version of Eq. (2), i.e., in the absence of external excitation. This requires, in general, the solution of $\left[1-\alpha_{\text{ee}}\breve{G}_{vv}^{\infty}\left(\mathbf{r}_{00},\mathbf{r}_{00},\mathbf{k}_{\text{B}}\right)\right]=0$ for complex $\mathbf{k}_{\text{B}}$. In what follows we let $\mathbf{k}_{\text{B}}=k_x^{\text{mode}}\hat{\mathbf{x}}$ and compute the free modes with complex wavenumber $k_x^{\text{mode}}=\beta_x+i\alpha_x$ ($\beta_x$ is the modal propagation constant and $\alpha_x$ the modal attenuation constant) for the three dipole polarizations $v=x,y,\text{or } z$. Due to symmetry both $\pm k_x^{\text{mode}}$ are free modal solutions that usually require classification as proper or improper; physical or nonphysical (i.e., excitable or non-excitable by a localized source); and forward or backward, as done in [25, 30-31]. However, here we are looking for modes classified as fast waves, i.e. $|\beta_x|<k$, that may be *forcibly excited* by a plane wave, and other explicit classifications as mentioned above are not relevant to the forced excitation of free modes and thus not stated in this paper. Moreover, only solutions with $\beta_x>0$ (with either positive or negative $\alpha_x$) are shown because we are interested in those modes that can be phase-matched to an oblique plane wave with positive, real transverse wavenumber $k_x^{\text{PW}}$, as we will see next. Interestingly, we observe that there are certain frequency ranges where the modes shown in Fig. 3 exhibit a small $\alpha_x$ and at the same time fall in the fast region, i.e., $\beta_x$ lies on the left of the light line $\beta_x=k$ (dashed grey curve), where $k=\omega\sqrt{\varepsilon_h}/c$ is the host wavenumber. The smaller $\alpha_x$, the more the plane wave will interact with the metasurface.

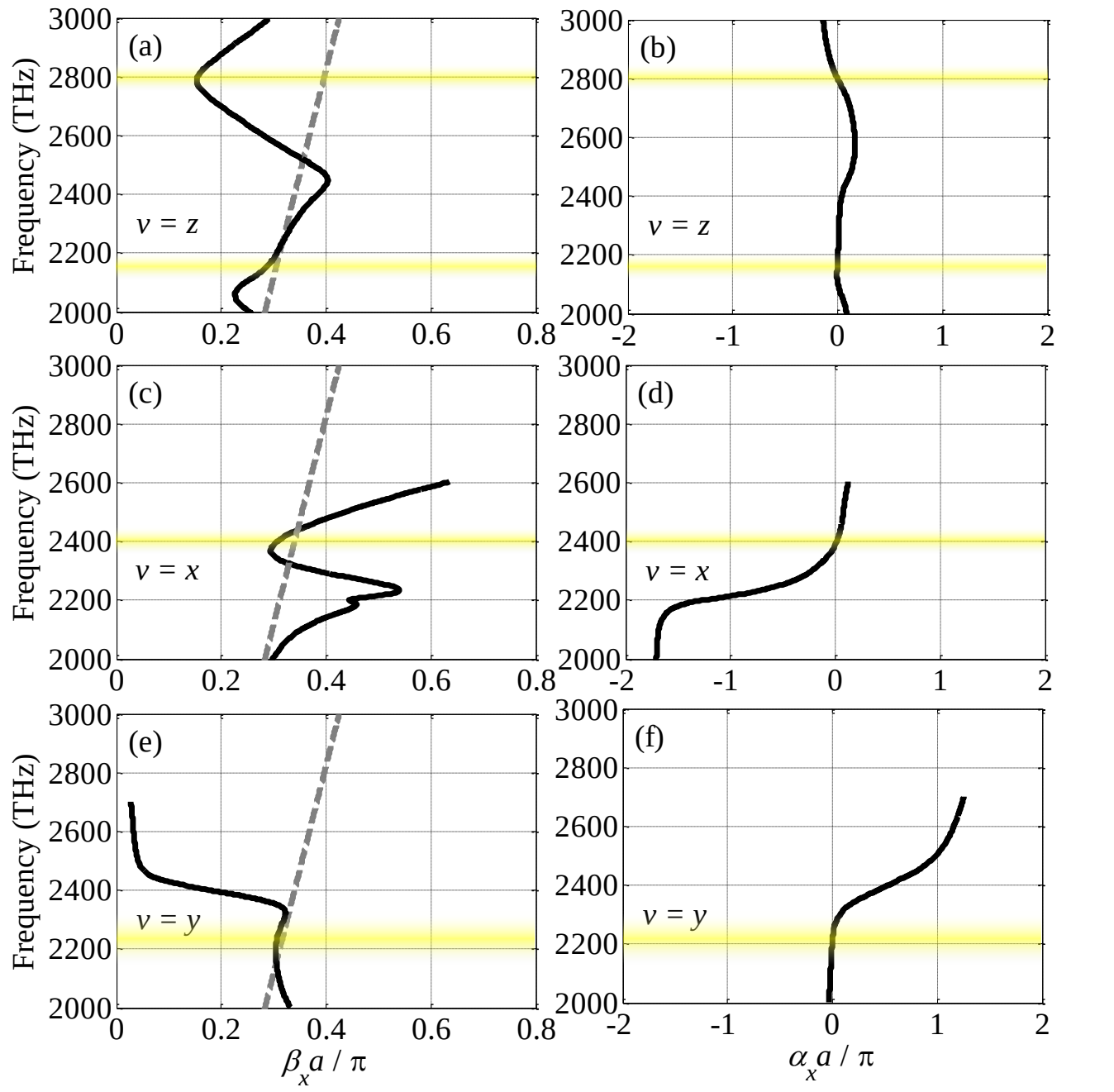

Fig. 3. (Color online) (a-f) Frequency-wavenumber dispersion diagrams for modes with Bloch wavenumber $\mathbf{k}_{\text{B}}=k_x^{\text{mode}}\hat{\mathbf{x}}$ and $x$, $y$, and $z$ polarization, respectively, relative to a structure as in Fig. 1 with $a=b=21.2\,\text{nm}$. The dashed grey line depicts the light line for which $\beta_x=k$. Modes with $\beta_x$ on the left side of the light line are leaky and modes on the right of the light line are bound to the array plane. Only leaky modes can be excited forcibly by an incoming plane wave. Highlighted regions denote fast wave bands with small attenuation constant.

The metasurface capability to support the free modes in Fig. 3 affects the array's optical properties when the array is excited by TE- or TM-polarized plane waves incident from the host medium, via phase-matching. We will see next that in the regions where the mode is leaky only *forced excitation* through a homogeneous plane wave may be verified. A *free* mode as in Fig. 3 would be perfectly matched to an external field when both real and imaginary parts of the transverse wavenumber $k_x^{\text{mode}}=\beta_x+i\alpha_x$ are matched to the complex wavenumber of the exciting field. Thus, a homogeneous plane wave with real $k_x^{\text{PW}}$ cannot excite a *free* mode, though *physical* free modes can be excited by the complex wavenumber spectrum generated by a finite-size source [21]. However, homogeneous plane waves may *force the excitation* of free modes (no matter if they are physical or unphysical [25, 32]) by phase-matching $k_x^{\text{PW}}$ to the real part $\beta_x$ of the complex wavenumber $k_x^{\text{mode}}$ when the imaginary part $\alpha_x$ of the modal wavenumber is sufficiently small. More properly speaking, the capability of supporting a mode with $\beta_x < k$ and $|\alpha_x| << k$ would affect the interaction of the metasurface with the incident plane wave if $\beta_x \approx k_x^{\mathrm{PW}}$.

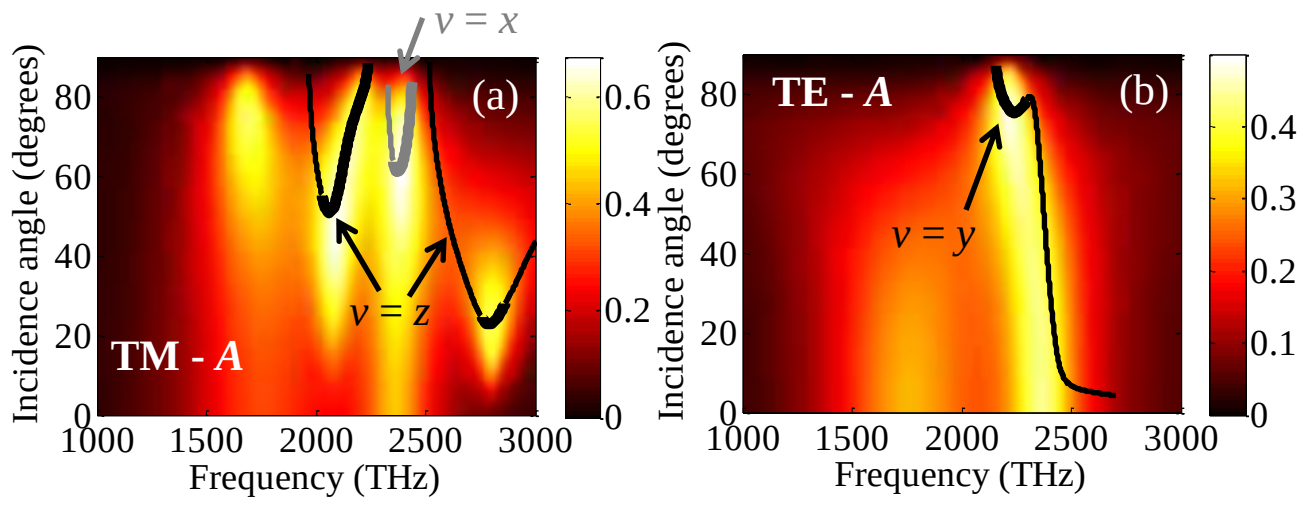

Fig. 4. (Color online) (a) Absorption versus frequency and incidence angle for a TM-polarized plane wave incidence, from SDA. In the absorption map, we have superimposed the modal wavenumber results from Fig. 3 relative to both *x*- (grey) and *z*-polarized (black) dipoles. Thick lines refer to modes with low attenuation constant, i.e., $|\alpha_x| < 0.07\pi / a$, thin lines otherwise. (b) As in (a), for a TE-polarized plane wave incidence. In the absorption map, we have superimposed the modal wavenumber result in Fig. 3 relative to *y*-polarized dipoles.

In order to verify these claims, in Fig. 4 we report absorption maps for TE- and TM-polarized incident plane waves versus incidence angle and frequency. In addition to the Mie resonances induced by the single nanoshell, the array-induced Fano resonances appear under oblique incidence due to the presence of the free Bloch modes shown in Fig. 3. This occurs when $\beta_x \approx k_x^{\mathrm{PW}}$, with $|\alpha_x| \ll k$. We now show that the absorption peaks seen under oblique incidence in Fig. 4 are due to the *forced excitation* of free modes supported by the array. The phase-matching angle for each of the modes in Fig. 3, i.e., the angle at which the real transverse wavenumber $k_x^{\mathrm{PW}}$ of the impinging plane wave matches the propagation constant of the modes, is compared to $\theta^{\mathrm{mode}} = \arcsin(\beta_x / k)$ given by the modal phase propagation constant. It is expected that *y*-polarized modes have the capability to interact with TE plane wave incidence, whereas, *x*- and *z*-polarized modes have the capability to interact with TM incident waves. Since the free modes supported by the metasurface exhibit quite different dispersion properties depending on the mode polarization, we expect that the metasurface spectral response for TE and TM incident wave will be dramatically different under oblique plane wave incidence, depending also on the capability of the wave to forcibly excite the metasurface modes. Angle-frequency dispersion curves associated with these modes are then plotted on the absorption maps versus frequency and incidence angle, showing good overlap when $|\alpha_x| a / \pi < 0.07$ (Fig. 4, thick curve). This condition presumably defines the only ranges where these modes may be forcibly excited by an incoming plane wave due to the low value of the attenuation constant $\alpha_x$. In particular, we observe that the curve pertaining to *y*-polarized dipoles correlates well with the narrow feature for TE-polarized plane wave incidence (black curve in Fig. 4, top row). Likewise, the curves pertaining to *x*- and *z*-polarized dipoles correlate with the several peaks observed for TM-polarized plane wave incidence (black and grey curves in Fig. 4, bottom row). This is an important aspect of metasurfaces: the array design and its modal phase and attenuation constants are central to the sharpness of the Fano resonances, e.g., see [15]. The narrow frequency regions highlighted in Fig. 3 coincide with the absorption peaks in Fig. 4.

This mode interaction is further justified by introducing an equivalent polarizability $\alpha_{\mathrm{ee},vv}^{\mathrm{array}}$ that *includes array effects* as [23]

$$p_{00} = \alpha_{\mathrm{ee},vv}^{\mathrm{array}} E_v^{\mathrm{inc}}(\mathbf{r}_{00})$$
$$\alpha_{\mathrm{ee},vv}^{\mathrm{array}} = \frac{\alpha_{\mathrm{ee}}}{\left[1 - \alpha_{\mathrm{ee}} \breve{G}_{vv}^{\infty}(\mathbf{r}_{00}, \mathbf{r}_{00}, \mathbf{k}_{\mathrm{B}})\right]}. \quad (3)$$

It is clear from Eq. (3) that array effects, accounted for by the term $\left[1 - \alpha_{\mathrm{ee}} \breve{G}_{vv}^{\infty}(\mathbf{r}_{00}, \mathbf{r}_{00}, \mathbf{k}_{\mathrm{B}})\right]$, can dramatically modify $\alpha_{\mathrm{ee},vv}^{\mathrm{array}}$, as reported in Fig. 5, and thus the optical properties of the array, for example the absorption shown in Fig. 2 and Fig. 4. Remember for a mode with complex $\mathbf{k}_{\mathrm{B}} = k_x^{\mathrm{mode}} \hat{\mathbf{x}}$, one has $\left[1 - \alpha_{\mathrm{ee}} \breve{G}_{vv}^{\infty}(\mathbf{r}_{00}, \mathbf{r}_{00}, \mathbf{k}_{\mathrm{B}})\right] = 0$, therefore the signature of a mode is seen in the array polarizability under plane wave incidence angles with $\mathbf{k}_{\mathrm{B}} = k_x^{\mathrm{PW}} \hat{\mathbf{x}}$ when $k_x^{\mathrm{PW}} \approx \mathrm{Re}\{k_x^{\mathrm{mode}}\}$ and $\mathrm{Im}\{k_x^{\mathrm{mode}}\}$ is very small. These regions are highlighted in Fig. 5. At normal incidence, there is no *z*-component of the electric field; thus, only the *xx* and *yy* components of $\alpha_{\mathrm{ee},vv}^{\mathrm{array}}$ will affect the optical properties of the metasurface. Resonances similar to Mie resonances supported by a single nanoshell shown in the inset in Fig. 1 are indeed observed at normal incidence in Fig. 5, proving polarization independence and absence of Fano resonances. By increasing the incidence angle, however, all the components of $\alpha_{\mathrm{ee},vv}^{\mathrm{array}}$ will affect the optical properties, depending on the polarization state: TE waves interact with the $\alpha_{\mathrm{ee},yy}^{\mathrm{array}}$ component, whereas TM waves interact with both the transverse ($\alpha_{\mathrm{ee},xx}^{\mathrm{array}}$) and the longitudinal ($\alpha_{\mathrm{ee},zz}^{\mathrm{array}}$) components (we recall that $\mathbf{k}_{\mathrm{B}}$ is along *x*). We observe that array polarizabilities are largely modulated in correspondence of absorption peaks observed in Fig. 4, further stressing the fact that array effects are the origin of the appearance of Fano resonances. In particular, at 30 degrees we observe four peaks in the $\alpha_{\mathrm{ee},xx}^{\mathrm{array}}$ and $\alpha_{\mathrm{ee},zz}^{\mathrm{array}}$ spectra (one at ~1750 THz and one at ~2400 THz for $\alpha_{\mathrm{ee},xx}^{\mathrm{array}}$; one at ~2100 THz and one at ~2800 THz for $\alpha_{\mathrm{ee},zz}^{\mathrm{array}}$) and two peaks in $\alpha_{\mathrm{ee},yy}^{\mathrm{array}}$ (one at ~1750 THz and one at ~2400 THz), in agreement with the four absorption peaks for TM incidence and the two absorption peaks for TE incidence observed in Fig. 2 and Fig. 4. Absorption peaks for increasing incidence angle can be explained in a similar manner looking at the corresponding $\alpha_{\mathrm{ee},vv}^{\mathrm{array}}$ spectra.

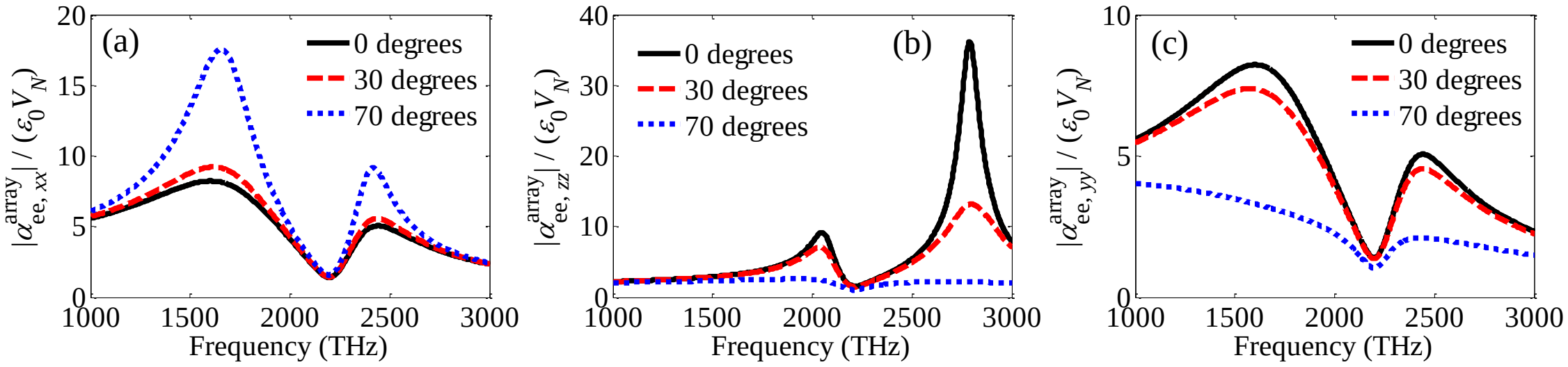

Fig. 5. (Color online) Magnitude of the equivalent array-polarizability $\alpha_{\mathrm{ee},\nu\nu}^{\mathrm{array}}$ for three incidence angles. (a) $\nu\nu = xx$ and (b) $\nu\nu = zz$ for TM polarization. (c) $\nu\nu = yy$ for TE polarization.

In conclusion, we have shown that the optical properties (like absorption) of 2D periodic arrays of plasmonic nanoshells and complex modes supported by these structures are intimately related. In particular, two kinds of resonances are observed in the array's optical properties: (i) dipolar Mie resonances, which are polarization and angle independent, and are related to the individual nanoshell polarizability reported in the inset in Fig. 1; and (ii) array-induced Fano resonances, due to the forced excitation of free modes supported by the structure (usually) under oblique incidence, which depend on polarization of the incident plane wave and dispersion of the free-modes' complex wavenumber. These array-induced resonances provide design flexibility to achieve enhanced, narrow optical properties in plasmonic metasurfaces by engineering the mode dispersion and the attenuation constant in particular. The use of modal analysis is then pivotal to properly explain the presence of Fano resonances at oblique incidence, for arbitrary incident polarization, and to provide an adequate, physical explanation of the resonant phenomena in metasurfaces composed of plasmonic nanoparticles. In turn, our findings show that modal analysis can indeed be a powerful design tool to develop filters and sensing devices, at ultraviolet frequencies for example, in addition to providing better understanding of wave propagation in periodic structures. Similar observations may be extended to other frequency ranges by proper material and geometrical choices.

## Acknowledgment

SC, CG and FC acknowledge partial support from the National Science Foundation under Grant No. CM.MI-1101074. This research was performed while the authors MAV and DdC held a National Research Council Research Associateship award at the U.S. Army Aviation and Missile Research Development and Engineering Center.

## References

[1] U. Fano, *Phys. Rev.* **124**, 1866-1878 (1961).
[2] Y. Francescato, V. Giannini, and S. A. Maier, *ACS Nano* **6**, 1830-1838 (2012).
[3] Z. Li, S. Butun, and K. Aydin, *ACS Nano*, 2014.
[4] W. Liu, A. E. Miroshnichenko, D. N. Neshev, and Y. S. Kivshar, *Phys. Rev. B* **86**, 081407 (2012).
[5] A. E. Miroshnichenko, S. Flach, and Y. S. Kivshar, *Rev. Mod. Phys.* **82**, 2257-2298 (2010).
[6] B. Luk'yanchuk, N. I. Zheludev, S. A. Maier, N. J. Halas, P. Nordlander, H. Giessen, and C. T. Chong, *Nat Mater* **9**, 707-715 (2010).
[7] J. A. Fan, C. Wu, K. Bao, J. Bao, R. Bardhan, N. J. Halas, V. N. Manoharan, P. Nordlander, G. Shvets, and F. Capasso, *Science* **328**, 1135-1138 (2010).
[8] M. Hentschel, D. Dregely, R. Vogelgesang, H. Giessen, and N. Liu, *ACS Nano* **5**, 2042-2050 (2011).
[9] J. Ye, F. Wen, H. Sobhani, J. B. Lassiter, P. V. Dorpe, P. Nordlander, and N. J. Halas, *Nano Lett.* **12**, 1660-1667 (2012).
[10] C. Wu, A. B. Khanikaev, R. Adato, N. Arju, A. A. Yanik, H. Altug, and G. Shvets, *Nat Mater* **11**, 69-75 (2012).
[11] V. A. Tamma, Y. Cui, J. Zhou, and W. Park, *Nanoscale* **5**, 1592-1602 (2013).
[12] B. S. Luk'yanchuk, A. E. Miroshnichenko, and S. K. Yu, *J. Opt.* **15**, 073001 (2013).
[13] C. Argyropoulos, F. Monticone, G. D'Aguanno, and A. Alù, *Appl. Phys. Lett.* **103**, 143113 (2013).
[14] S. Campione, C. Guclu, R. Ragan, and F. Capolino, *Opt. Lett.* **38**, 5216-5219 (2013).
[15] S. Campione, C. Guclu, R. Ragan, and F. Capolino, *ACS Photonics* **1**, 254-260 (2014).
[16] S. H. Mousavi, A. B. Khanikaev, and G. Shvets, *Phys. Rev. B* **85**, 155429 (2012).
[17] V. G. Kravets, F. Schedin, and A. N. Grigorenko, *Phys. Rev. Lett.* **101**, 087403 (2008).
[18] A. Vallecchi, M. Albani, and F. Capolino, *Opt. Express* **19**, 2754-2772 (2011).
[19] F. Shafiei, F. Monticone, K. Q. Le, X.-X. Liu, T. Hartsfield, A. Alu, and X. Li, *Nat Nano* **8**, 95-99 (2013).
[20] G. D'Aguanno, N. Mattiucci, M. J. Bloemer, R. Trimm, N. Aközbek, and A. Alù, *Phys. Rev. B* **90**, 054202 (2014).
[21] A. Hessel and A. A. Oliner, *Appl. Opt.* **4**, 1275-1297 (1965).
[22] D. de Ceglia, S. Campione, M. A. Vincenti, F. Capolino, and M. Scalora, *Phys. Rev. B* **87**, 155140 (2013).
[23] B. Auguié and W. L. Barnes, *Phys. Rev. Lett.* **101**, 143902 (2008).
[24] S. Steshenko and F. Capolino, "Single Dipole Approximation for Modeling Collection of Nanoscatterers," in *Theory and Phenomena of Metamaterials*, F. Capolino, Ed., ed Boca Raton, FL: CRC Press, 2009, p. 8.1.
[25] A. L. Fructos, S. Campione, F. Capolino, and F. Mesa, *J. Opt. Soc. Am. B* **28**, 1446-1458 (2011).
[26] S. Steshenko, F. Capolino, P. Alitalo, and S. Tretyakov, *Phys. Rev. E* **84**, 016607 (2011).
[27] E. Palik, *Handbook of Optical Constants of Solids*. New York: Academic Press, 1985.
[28] C. F. Bohren and D. R. Huffman, *Absorption and Scattering of Light by Small Particles*. New York: Wiley, 1983.
[29] S. Campione, S. Pan, S. A. Hosseini, C. Guclu, and F. Capolino, "Electromagnetic metamaterials as artificial composite structures, Third Edition," in *Handbook of Nanoscience, Engineering, and Technology*, B. Goddard, *et al.*, Eds., 3rd ed Boca Raton, FL: CRC Press, 2012.
[30] P. Baccarelli, S. Paulotto, and C. D. Nallo, *IET Microw. Antennas Propagat.* **1**, 217-225 (2007).
[31] F. Capolino, D. R. Jackson, and D. R. Wilton, "Field representations in periodic artificial materials excited by a source," in *Theory and Phenomena of Metamaterials*, F. Capolino, Ed., ed Boca Raton, FL: CRC Press, 2009, p. 12.1.

[32] S. Campione, S. Steshenko, and F. Capolino, *Opt. Express* **19**, 18345-18363 (2011).